\documentclass[runningheads,a4paper]{llncs}

\usepackage[numbers]{natbib} 
\makeatletter
\renewcommand\bibsection%
{
  \section*{\refname
    \@mkboth{\MakeUppercase{\refname}}{\MakeUppercase{\refname}}}
} \makeatother

\usepackage{graphicx}
\usepackage{amssymb} 
\usepackage[misc]{ifsym} 
\usepackage[labelsep=colon,margin=10pt,labelfont=bf]{caption} 
\usepackage[english]{babel}
\usepackage[latin1]{inputenc}
\usepackage[active]{srcltx}
\usepackage{url}
\usepackage{amsmath}
\usepackage{verbatim}
\usepackage{array}    
\usepackage{tabularx} 
\usepackage{algorithmic}
\makeatletter
\newif\if@restonecol
\makeatother

\usepackage[boxed]{algorithm2e}  
\usepackage{colortbl} 

\urldef{\mailsa}\path|{sergio.consoli@jrc.it,|
\urldef{\mailsb}\path|jamoreno@ull.es,| 
\urldef{\mailsc}\path|nenad.mladenovic@brunel.ac.uk}|

\begin{document}

\mainmatter  

\title{Towards an intelligent VNS heuristic 
for the \emph{k}-labelled spanning forest problem}

\titlerunning{Intelligent VNS 
for the kLSF problem}

%

\author{Sergio Consoli\inst{1}%
\and Jos\'{e} Andr\'{e}s Moreno P\'{e}rez\inst{2} \and Nenad
Mladenovi\'{c}\inst{3}}

\authorrunning{Consoli et al.}

\institute{ISTC/STLab, National Research Council (CNR), Catania, Italy \and
Department of Computing Engineering, Universidad de La Laguna, Tenerife, Spain \and School of Computing \& Mathematics, Brunel University, London, United Kingdom \mailsa
\mailsb 
\mailsc }

\toctitle{Intelligent VNS
for the kLSF problem} \tocauthor{Consoli et al.}
\maketitle


In a currently ongoing project, we investigate a new possibility for solving the \emph{k}-labelled spanning forest (kLSF) problem by an intelligent Variable Neighbourhood Search (Int-VNS) metaheuristic. In the kLSF problem we are given an undirected input graph $G=(V,E,L)$ where
$V$ is the set of nodes, $E$ the set of edges, that are labelled
on the set $L$ of labels, 
and an integer positive value $\bar{k}$, and the aim is to find a spanning forest $G^*=(V,E^*,L^*)$ of the input graph having the minimum number of connected components, i.e. $min |Comp(G^*)|$, and the upper bound $\bar{k}$ on the number of labels to use, $|L^*| \leq \bar{k}$. The problem is related to the minimum labelling spanning tree (MLST) problem~\citep{Chang}, whose goal is to get the spanning tree of the input graph with the minimum number of labels, and has several applications in the real-world, 
where one aims to ensure connectivity by means of homogeneous connections. 
The kLSF 
problem was recently introduced in~\citep{CerulliKLSF} along with the proof of its NP-hardness; therefore any practical solution approach 
requires heuristics~\cite{CerulliKLSF,VNSkLSF2014}. In particular our aim is to present
an intelligent VNS which is aimed to achieve further improvements for the kLSF problem. This approach is derived from the promising strategy recently proposed in~\citep{MLSTP-int-INOC} for the MLST problem, and
integrates the basic VNS for the kLSF problem in~\citep{VNSkLSF2014} with other complementary approaches from machine learning, statistics and experimental algorithmics, 
in order to produce high-quality performance and to completely automate the resulting strategy.

The first extension that we introduce 
is a 
local search mechanism that is inserted at top of the basic VNS. 
The resulting
local search method is referred to as~\emph{Complementary Variable
Neighbourhood Search} (Co-VNS)~\citep{MLSTP-int-INOC}. 
Given a labelled graph $G =
(V,E,L)$ with $n$ vertices, $m$ edges, and $\ell$ labels, Co-VNS
replaces iteratively each incumbent solution $L^* = (l_1, l_2 ,..., l_{\ell})$, where, $\forall i=1,\ldots,|L|$, $\l_i = 1$ if label $i
\in L^*$, $\l_i = 0$ otherwise, with another 
solution selected from the \emph{complementary space} of $L^*$, referred to as $Co_{space}$, defined as the set of all the labels that are not contained in $L^*$, that is $L\Delta L^*$. 
The iterative process of extraction 
of a new 
solution from the complementary space of the
current solution 
helps to escape the algorithm from possible traps 
in local minima, since the complementary solution lies 
in a very different zone of the search space with respect to the
incumbent solution. 
Successively, the basic VNS~\citep{VNSkLSF2014} 
is applied in order to improve the resulting solution. At the
starting point of VNS, it is required to define a suitable
neighbourhood structure of size $q_{max}$. 
The simplest and most common choice is a structure in which the
neighbourhoods have increasing cardinality: $|N_1(\cdot)| <
|N_2(\cdot)| < ... < |N_{q_{max}}(\cdot)|$. In order to impose a
neighbourhood structure on the solution space $S$, comprising all
possible solutions, we define the distance between any two such
solutions $L_1, L_2 \in S$, as the Hamming distance: $\rho
(L_1,L_2)=|L_1 \Delta L_2|=\sum_{i=1}^{\ell}\lambda_i$,
where $\lambda_i = 1$ if label $i$ is included in one of the
solutions but not in the other, and $0$ otherwise, $\forall i =
1,..., \ell$. 
In order to construct the neighbourhood of a solution $L^*$, the
algorithm first proceeds with the deletion of labels from $L^*$. In
other words, given a solution $L^*$, its $q^{th}$ neighbourhood,
$N_q(C)$, consists of all the different sets obtained from $L^*$ by
removing $q$ labels, where $q \leftarrow 1,2,..., q_{max}$. 
In order to seek further improvements and to automate
on-line the search process, Co-VNS has been modified by including a probability-based local
search with a self-tuning parameters setting, resulting in the intelligent VNS metaheuristic that we propose. 
The probability-based local search is 
obtained by introducing a probabilistic
choice on the next label to be added into incomplete solutions. By
allowing worse components to be added to incomplete solutions, this
probabilistic constructive heuristic produces a further increase on the diversification of the optimization process. The construction
criterion is as follows. The procedure starts from an initial
solution and iteratively selects at random a candidate move. If this
move leads to a solution having a better objective function value
than the current solution, then this move is accepted
unconditionally; otherwise the move is accepted with a probability
that depends on the deterioration, $\Delta$, of the objective
function value.
This construction criterion takes inspiration from Simulated
Annealing: 
the acceptance probability of a
worse component into a partial solution is evaluated according to
the Boltzmann function $\exp(-\Delta /T)$,
where the parameter $T$, referred to as \emph{temperature}, controls
the dynamics of the function. Initially the value of $T$ is large,
so allowing many worse moves to be accepted, and is gradually
reduced by the following geometric cooling law:
$T_{j+1} = \alpha \cdot T_j$, where $T_{0}=|Best_L|$ and $\alpha=1/|Best_L| \in [0, 1]$, with $Best_L$ being the current best solution and $|Best_L|$ its
number of labels. 
This cooling schedule
does not requires any intervention from the user regarding the
setting of its parameters, as it is guided automatically by the best
solution $Best_L$. Therefore the whole process is able
to react in response to the search algorithm's behavior and to adapt
its setting on-line according to the instance of the problem under
evaluation. 

The achieved optimization strategy seems to be highly promising for the kLSF problem. Ongoing 
investigation will consist in a statistical comparison of the
resulting strategy against the best kLSF algorithms in the
literature, in order to quantify and qualify the improvements
obtained by the proposed Int-VNS.



\bibliographystyle{splncs}
\begin{scriptsize}
\bibliography{bibliography}
\end{scriptsize}

\end{document}